\documentclass{vgtc} 

\usepackage{mathptmx}
\usepackage{graphicx}
\usepackage{times}
\usepackage{amsmath}

\usepackage[utf8]{inputenc}
\usepackage[normalsize]{subfigure}

\usepackage{cite} 
\usepackage[final]{hyperref} 
\hypersetup{
	colorlinks=true,       
	linkcolor=blue,        
	citecolor=blue,        
	filecolor=magenta,     
	urlcolor=blue         
}

\onlineid{0}

\title{Power coupling losses for misaligned and mode-mismatched higher-order Hermite-Gauss modes}

\author{Liu Tao$^{1}$\thanks{Corresponding author: liu.tao@ligo.org} %
\and Jessica Kelley-Derzon$^{2}$ %
\and Anna C. Green$^{1}$ %
\and Paul Fulda$^{1}$}
\affiliation{\scriptsize  $^{1}$University of Florida, 2001 Museum Road, Gainesville, Florida 32611, USA \\
 $^{2}$Skidmore College, Saratoga Springs NY 12866, USA
}

\abstract{This paper analytically and numerically investigates  misalignment and mode-mismatch induced power coupling coefficients and losses as a function of Hermite Gauss (HG) mode order. We show that higher-order HG modes are more susceptible to beam perturbations when, for example, 
coupling into optical cavities: the misalignment and mode-mismatch-induced power coupling losses scale linearly and quadratically with respect to the mode indices respectively. As a result, the mode-mismatch tolerance for the $\mathrm{HG}_{3,3}$ mode is reduced to a factor of 0.28 relative to the currently-used $\mathrm{HG}_{0,0}$ mode. This is a potential hurdle to using higher-order modes to reduce thermal noise in future gravitational-wave detectors.} 

\begin{document}

\maketitle
\section{Introduction}
There is increasing interest in replacing the fundamental Gaussian laser beam used in all current gravitational-wave detectors~\cite{aLIGO,AdVirgo} with beams of more uniform intensity distribution, such as higher-order Hermite-Gauss (HG) modes. This would reduce the thermal noise of the test-mass optics~\cite{ Mours_2006, Vinet_2007}, which limits detector sensitivity at signal frequencies around 100 Hz. It has been shown that higher-order HG modes such as the $\mathrm{HG_{3,3}}$ mode are nearly as robust against  mirror surface deformations as the fundamental $\mathrm{HG_{0,0}}$ mode when vertical astigmatism is deliberately added to the test-mass optics~\cite{PhysRevD.102.122002, PhysRevD.103.042008}.
However A. Jones \emph{et. al.}~\cite{Jones_2020} have shown, using the computational algebra system SymPy~\cite{sympy}, that the mode-mismatch-induced power losses increase monotonically with mode index when, for example,
coupling into optical cavities. This paper takes an analytical approach, and extends their work to include the case of misalignment. 

Sec.~\ref{sec:2} derives the mode content and resulting misalignment and mode-mismatch induced power coupling coefficients and losses for arbitrary higher-order HG modes by Taylor expanding the beam spatial profile functions up to second order in the perturbation under consideration. Sec.~\ref{sec:3} then uses a numerical approach, representing the original and perturbed beams as discrete matrices. This shows good agreement with the
analytical results. We report our conclusions and discussions in Sec.~\ref{sec:4}.

\section{Analytical calculations}
\label{sec:2}
A beam perturbed from a state considered to be an eigenmode of a basis, such as the HG mode basis, can be described as a mixture of the original mode, $\Psi_{0}$ and other  eigenmodes into which power is `scattered'.
The lowest order perturbations of importance are misalignment and mode mismatch.

This scattering effect is characterized by
the overlap between the perturbed beam, $\Psi^{\prime}$ and the original mode, known as the \textit{mode coupling coefficient}~\cite{Bayer-Helms:84}:
\begin{equation}
\rho \equiv \iint_{-\infty}^{\infty} \Psi^{\prime} \Psi_{0}^{\ast}\,dxdy \,.
\end{equation}
In general $\rho$ is complex; in our case it is more useful to consider
the scattering effect in terms of the real-valued \textit{power coupling coefficient}:
\begin{equation}
\eta \equiv \rho \cdot \rho^{\ast} \,.
\label{equ:genpowcoeff}
\end{equation}
which we will calculate up to the second order in its Taylor series expansion in this manuscript. 
We also define the induced \textit{relative power coupling loss}, 
which for convenient comparison is 
normalized by the result for the fundamental Gaussian mode:
\begin{equation}
\Gamma \equiv \frac{1 - \eta}{1 - \eta_{0}} \,,
\label{equ:powerloss}
\end{equation}
A larger value of $\Gamma$ indicates a system with a lower tolerance for a given beam perturbation such as the misalignment or mode mismatch, normalized by the tolerance for the fundamental mode.

In this section we analytically derive 
the misalignment and mode-mismatch induced power coupling coefficients and losses for arbitrary higher-order $\mathrm{HG}_{\mathrm{n,m}}$ modes propagating along the $z$ axis. The general expression for a Hermite-Gauss mode is
\cite{Bond2017}
\begin{equation}
\mathcal{U}_{\mathrm{nm}}(x, y, z)=\mathcal{U}_{\mathrm{n}}(x, z) \mathcal{U}_{\mathrm{m}}(y, z)
\label{equ:HGnm}
\end{equation}
with 
\begin{equation}
\begin{aligned}
\mathcal{U}_{\mathrm{n}}(x,y, z)&=\left(\frac{2}{\pi}\right)^{1 / 4}\left(\frac{\exp (\mathrm{i}(2 n+1) \Psi(z))}{2^{n} n ! w(z)}\right)^{1 / 2} \\
& \times H_{n}\left(\frac{\sqrt{2} x}{w(z)}\right) \exp \left(-\mathrm{i} \frac{k x^{2}}{2 R_{c}(z)}-\frac{x^{2}}{w^{2}(z)}\right) \,,
\end{aligned}
\label{equ:HGn0}
\end{equation}
where $\Psi(z) = \arctan \left(\frac{z-z_{0}}{z_{R}}\right)$ is the Gouy phase with $z_{R} = \frac{\pi w_{0}^2}{\lambda}$ being the Rayleigh range. $k$ is the wavenumber, $\lambda$ is the wavelength, $w(z)$ is the beam radius and $R_{c}(z)$ is the wavefront radius of curvature. 

In the following we use two properties of the Hermite polynomials $H_{n}(\frac{\sqrt{2} x}{w_{0}})$:
\begin{alignat}{2}
 2 \frac{\sqrt{2} x}{w_{0}} H_{n} &= H_{n+1} + 2n H_{n-1} \label{equ:hermite11}\\  
 H^{\prime}_{n} &= 2n H_{n-1} \,.
\label{equ:hermite12}
\end{alignat}
The function argument $\frac{\sqrt{2}x}{w_{0}}$ is implied throughout this manuscript and the derivative is applied with respect to this argument. Applying Eq.~\ref{equ:hermite11} twice, we can write:
\begin{equation}
\begin{aligned}
\frac{x^{2}}{w_{0}^{2}}H_{n} = \frac{1}{8} \Big( H_{n+2} + 2(2n+1)H_{n} + 4n(n-1)H_{n-2}\Big)\,,
\end{aligned}
\label{equ:hermite2}
\end{equation}

and applying four times:
\begin{equation}
\begin{aligned}
&\frac{x^{4}}{w_{0}^{4}}H_{n} = \frac{1}{64}\Big(H_{n+4} + 4(2n+3)H_{n+2} + 12(2n^2+2n+1)H_{n} \\
&+ 16n(n-1)(2n-1)H_{n-2} + 16n(n-1)(n-2)(n-3) H_{n-4}\Big)\,.
\end{aligned}
\label{equ:hermite3}
\end{equation}
The relation shown in Eq.~\ref{equ:hermite12} can also be used twice to write
\begin{equation}
H^{\prime \prime}_{n} = 4n(n-1) H_{n-2}\,.
\end{equation}

\subsection{Misalignment}
\label{sec:misalign}
HG modes are separable in $x$ and $y$, so for misalignment we can consider the single-axis behaviour without loss of generality. We therefore consider a $\mathrm{HG}_{\mathrm{n},0}$ mode propagating along the $z$ axis and explore the effect of misalignment in the $x$-$z$ plane. 

\subsubsection{Misalignment: tilt}
Any small misalignment can be resolved into a combination of a lateral displacement and a tilt at the beam waist. First we consider a tilt about the waist in the $x$-$z$ plane, $\alpha$, between the perturbed beam axis and the unperturbed optical axis.
The tilted beam can be described in the original basis as having an additional transverse phase term. 
For small angles ($\alpha \ll \Theta$), this can be Taylor-expanded to second order as:
\begin{equation}
\exp \left(\mathrm{i}\frac{2\pi \alpha}{\lambda} x\right) = \exp \left(\mathrm{i}\frac{2 \alpha}{\Theta} \frac{x}{w_{0}}\right) \approx 1+\mathrm{i} \frac{2 \alpha}{\Theta} \frac{x}{w_{0}} - \frac{2 \alpha}{\Theta} \frac{x^2}{w_{0}^2} \,,
\end{equation}
where $\Theta = \frac{\lambda}{\pi w_{0}}$ is the far-field divergence angle, and $w_{0}$ is the beam waist size. At the beam waist $w(z)=w_{0}$ and $R_{c}(z)=\infty$, so the tilted input beam (Eq.~\ref{equ:HGn0}) in this approximation becomes
\begin{equation}
\begin{aligned}
\mathcal{U}_{n}^\mathrm{tilt}(x, z)&=\left(\frac{2}{\pi}\right)^{1 / 4}\left(\frac{\exp (\mathrm{i}(2 n+1) \Psi)}{2^{n} n ! w_{0}}\right)^{1 / 2} H_{n}\left(\frac{\sqrt{2} x}{w_{0}}\right) \\
&\times \exp \left(-\frac{x^{2}}{w_{0}^{2}}\right) \left(1+\mathrm{i} \frac{2 \alpha}{\Theta} \frac{x}{w_{0}} - \frac{2\alpha^{2}}{\Theta^{2}} \frac{x^2}{w_{0}^{2}}\right)\,.
\end{aligned}
\end{equation}
Then using Eqs.~\ref{equ:hermite11} and~\ref{equ:hermite2} we find 
\begin{equation}
\begin{aligned}
\mathcal{U}_{n}^\mathrm{tilt}(x, z)&= \mathcal{U}_{\mathrm{n}}(x)
+ \mathrm{i} \frac{\alpha}{\Theta} \Big(\sqrt{n+1} \mathcal{U}_{\mathrm{n}+1}e^{- \mathrm{i}\Psi} + \sqrt{n} \mathcal{U}_{\mathrm{n}-1} e^{\mathrm{i}\Psi}\Big) \\
&- \frac{\alpha^2}{2\Theta^{2}} \big( \sqrt{(n+1)(n+2)}\mathcal{U}_{n+2}e^{-2\mathrm{i}\Psi} + (2n+1) \mathcal{U}_{n} \\
&+ \sqrt{n(n-1)}\mathcal{U}_{n-2}e^{ 2\mathrm{i}\Psi}\big)\,.
\end{aligned}
\end{equation}
To first order, we see that tilt scatters $\mathrm{HG_{n, 0}}$ into $\mathrm{HG_{n\pm1, 0}}$. 

The mode coupling coefficient is
\begin{equation}
\rho = \int_{-\infty}^{\infty} dx \cdot \mathcal{U}_{n}^\mathrm{tilt} \cdot \mathcal{U}_{\mathrm{n}}^{\ast}  \approx 1 - \frac{\alpha^2}{2\Theta^{2}} \left(2n+1\right)\,,
\end{equation}
so the power coupling coefficient (Eq.~\ref{equ:genpowcoeff}) for $\mathrm{HG_{n,0}}$ due to tilt $\alpha$ is
\begin{equation}
\eta^\mathrm{tilt} \approx  1 - \frac{\alpha^2}{\Theta^{2}} \left(2n+1\right)
\label{equ:10}
\end{equation}
to second order. The relative power coupling loss (Eq.~\ref{equ:powerloss}) $\Gamma_{n}^\mathrm{tilt}$ becomes
\begin{equation}
\Gamma_{n}^\mathrm{tilt} = 2n+1\,,
\label{equ:tiltloss}
\end{equation}
where $n$ is the mode index of the beam. We thus see the relative power coupling loss for $\mathrm{HG_{n,0}}$ as a result of tilt between the beam axis and the unperturbed optical axis scales linearly with mode order. A simple propagation of the beam does not scatter power between modes so this result must be valid for all $z$-axis positions, not just the waist location.

\subsubsection{Misalignment: lateral offset}
For a small lateral displacement $\delta x_{0}\ll w_{0}$ along the $x$ direction, the displaced beam (Eq.~\ref{equ:HGn0}) can be Taylor-expanded at the waist to second order:
\begin{equation}
\begin{aligned}
\mathcal{U}&^\mathrm{offset}(x, z)=\left(\frac{2}{\pi}\right)^{1 / 4}\left(\frac{e^{\mathrm{i}(2 n+1) \Psi}}{2^{n} n ! w_{0}}\right)^{1 / 2} H_{n}\left(\frac{\sqrt{2} (x-\delta x_{0})}{w_{0}}\right) e^{-\frac{(x-\delta x_{0})^{2}}{w_{0}^{2}}} \\
& \approx \left(\frac{2}{\pi}\right)^{1 / 4}\left(\frac{e^{\mathrm{i}(2 n+1) \Psi}}{2^{n} n ! w_{0}}\right)^{1 / 2} \Big(H_{n} - \frac{\sqrt{2} \delta x_{0}}{w_{0}} 2n H_{n-1} + \frac{\delta x_{0}^{2}}{w_{0}^{2}} \\
&\times 4n(n-1) H_{n-2}\Big)  e^{-\frac{x^{2}}{w_{0}^{2}}}
\Big(1+2\frac{\delta x_{0}}{w_{0}^{2}} x - \frac{ \delta x_{0}^{2}}{w_{0}^2} + \frac{2 \delta x_{0}^{2} x^{2}}{w_{0}^{4}}\Big)\,.
\end{aligned}
\end{equation}
Using identities~\ref{equ:hermite11} and~\ref{equ:hermite2} and simplifying yields
\begin{equation}
\begin{aligned}
\mathcal{U}^\mathrm{offset}(x, z) &\approx \mathcal{U}_{n} + \frac{\delta x_{0}}{w_{0}} \left( \sqrt{n+1}\mathcal{U}_{n+1}e^{-\mathrm{i}\Psi} - \sqrt{n}\mathcal{U}_{n-1}e^{\mathrm{i}\Psi}\right) + \frac{ \delta x_{0}^{2}}{2w_{0}^2}\\
&\times \Big(-(2n+1) \mathcal{U}_{n} + \sqrt{(n+1)(n+2)}\mathcal{U}_{n+2}e^{-2\mathrm{i}\Psi} \\
&+\sqrt{n(n-1)}\mathcal{U}_{n-2}e^{2\mathrm{i}\Psi} \Big)\,.
\end{aligned}
\end{equation}
Collecting the coefficients of $\mathcal{U}_{n}$, the mode coupling coefficient is 
\begin{equation}
\begin{aligned}
\rho =  \int_{-\infty}^{\infty} dx \cdot \mathcal{U}^\mathrm{offset} \cdot \mathcal{U}_{\mathrm{n}}^{\ast} \approx 1 - \frac{ \delta x_{0}^{2}}{2 w_{0}^{2}} \left(2n+1\right)
\end{aligned}
\end{equation}
and the power coupling coefficient, to second order, is therefore
\begin{equation}
\eta^\mathrm{offset} \approx  1 - \left(2n+1\right)\frac{ \delta x_{0}^{2}}{w_{0}^{2}}\,.
\end{equation}
In this case the relative power coupling loss also scales linearly with respect to the mode order:
\begin{equation}
\Gamma_{n}^\mathrm{offset} = 2n+1 \,,
\label{equ:offsetloss}
\end{equation}
which equals $\Gamma_{n}^\mathrm{tilt}$ (Eq.~\ref{equ:tiltloss}).

\subsection{Mode mismatch}
Mode mismatches cannot be reduced to a single-axis treatment. Therefore we
consider a generic $\mathrm{HG_{n,m}}$ 
beam represented by the transverse function $\mathcal{U}_{\mathrm{n},\mathrm{m}}(x, y, z)$  
as defined in Eq.~\ref{equ:HGnm}. As in section~\ref{sec:misalign}, we consider the effect of perturbations at the cavity waist.

\subsubsection{Mode mismatch: waist position mismatch}
For a beam waist displacement $\delta z_{0}$ along the $z$-direction, the wavefront radius of curvature $R_{c}$ of the input beam at the cavity waist is no longer infinite.
Assuming a small displacement such that $\frac{\lambda \delta z_{0}}{\pi w_{0}^{2}} \ll 1$, this can be approximated as~\cite{Morrison1:94}
\begin{equation}
\begin{aligned}
\frac{1}{R_{c}}\approx -\left(\frac{\lambda}{\pi w_{0}^{2}}\right)^{2} \cdot \delta z_{0} \,.
\end{aligned}
\label{equ:curvaturemismatch}
\end{equation}

As a result, $\mathcal{U}_{\mathrm{n}}(x,z)$ becomes
\begin{equation}
\begin{aligned}
\mathcal{U}_{n}^{WP}(x, z)  \approx & \left(\frac{2}{\pi}\right)^{1 / 4}\left(\frac{\exp (\mathrm{i}(2 n+1) \Psi(z))}{2^{n} n ! w(z)}\right)^{1 / 2}  H_{n}\left(\frac{\sqrt{2} x}{w(z)}\right) \\
&\times 
e^{-\frac{x^{2}}{w^{2}(z)}}
\Big(1+\mathrm{i} \frac{ \lambda \delta z_{0}}{\pi w_{0}^{2}} \frac{x^{2}}{w_{0}^{2}} - \frac{ \lambda^2 \delta z_{0}^2}{2 \pi^2 w_{0}^{4}} \frac{x^{4}}{w_{0}^{4}} \Big) \,.
\label{equ:WPmm1}
\end{aligned}
\end{equation}

Applying Eqs.~\ref{equ:hermite11} and~\ref{equ:hermite2}, and writing $\gamma= \frac{kw_{0}^2}{R_{c}} \approx -\frac{2\lambda}{\pi w_{0}^2} \delta z$ for convenience, Eq.~\ref{equ:WPmm1} becomes
\begin{equation}
\begin{aligned}
\mathcal{U}&_{n}^{WP}(x, z) \approx \mathcal{U}_{n} -\mathrm{i} \frac{\gamma}{8} \Big(\sqrt{(n+1)(n+2)}\cdot \mathcal{U}_{n+2}e^{ -2\mathrm{i}\Psi} + (2n+1)\cdot \mathcal{U}_{n} \\
&+ \sqrt{n(n-1)}\cdot \mathcal{U}_{\mathrm{n-2}} e^{ 2\mathrm{i}\Psi} \Big) -\frac{\gamma^2}{128} \bigg(\sqrt{(n+1)(n+2)(n+3)(n+4)}\\
&\times \mathcal{U}_{\mathrm{n+4}} e^{ -4\mathrm{i}\Psi} + 2(2n+3)\sqrt{(n+1)(n+2)}\mathcal{U}_{\mathrm{n+2}} e^{ -2\mathrm{i}\Psi} \\
&+ 3(2n^2+2n+1)\mathcal{U}_{\mathrm{n}} + 2(2n-1)\sqrt{n(n-1)} \mathcal{U}_{\mathrm{n-2}}e^{ 2\mathrm{i}\Psi} \\
&+ \sqrt{n(n-1)(n-2)(n-3)}\mathcal{U}_{\mathrm{n-4}}e^{ 4\mathrm{i}\Psi}\bigg)\,.
\end{aligned}
\end{equation}
To first order, we see that waist position mismatch scatters  $\mathrm{HG_{n, 0}}$ into $\mathrm{HG_{n\pm2, 0}}$.

The mode coupling coefficient in the $x$-direction is 
\begin{equation}
\begin{aligned}
\rho_{x} &= \int_{-\infty}^{\infty} dx \cdot \mathcal{U}_{n}^{WP} \cdot \mathcal{U}_{\mathrm{n}}^{\ast} \\
&\approx 1- \mathrm{i}\frac{2 n+1}{8}\cdot \gamma -\frac{3\left(2n^2+2n+1\right)}{128} \cdot \gamma^2 
\end{aligned}
\label{equ:linear}
\end{equation}


We will have a similar result for the coupling coefficient in $y$, $\rho_{y}$, so the full mode coupling coefficient due to waist position mismatch is
\begin{equation}
\begin{aligned}
\rho = \rho_x \cdot \rho_y &\approx  1  - \mathrm{i}\frac{ n+m+1}{4} \cdot \gamma \\
&- \frac{ 3\left(n^2+m^2\right)+5\left(n+m\right)+4nm+4}{64} \cdot \gamma^2 
\end{aligned}
\end{equation}

The power coupling coefficient (Eq.~\ref{equ:genpowcoeff}) to second order is then
\begin{equation}
\eta^{WP} \approx 1-\frac{\left(n^2+n+1\right)+\left(m^2+m+1\right)}{32}\cdot \gamma^2.
\end{equation}
 Note that the linear term in $\eta^{WP}$ cancels out since $C_{n,m}$ is purely imaginary. The relative power coupling loss (Eq.~\ref{equ:powerloss}) due to waist position mismatch therefore scales quadratically with mode indices $n,m$:
\begin{equation}
\Gamma_{n,m}^{WP} = \frac{n^2+n+m^2+m+2}{2} \,.
\label{equ:WPm}
\end{equation}

\subsubsection{Mode mismatch: waist size mismatch}
In terms of the relative waist size mismatch parameter $\epsilon \equiv \frac{w}{w_{0}}-1$, $\mathcal{U}_{\mathrm{n}}$ can be written as
\begin{equation}
\begin{aligned}
&\mathcal{U}_{\mathrm{n}}^{WS}(x, z)  = \left(\frac{2}{\pi}\right)^{\frac{1}{4}}\left(\frac{e^{\mathrm{i}(2 n+1) \Psi(z)}}{2^{n} n ! w_{0}(1+\epsilon)}\right)^{\frac{1}{2}} H_{n}\left(\frac{\sqrt{2} x}{w_{0}(1+\epsilon)}\right) e^{-\frac{x^{2}}{(w_{0}(1+\epsilon))^{2}}} \\
& \approx \left(\frac{2}{\pi}\right)^{\frac{1}{4}}\left(\frac{\exp (\mathrm{i}(2 n+1) \Psi(z))}{2^{n} n ! w_{0}}\right)^{\frac{1}{2}} \left(1 - \frac{\epsilon}{2} + \frac{3}{8}\epsilon^2\right) \Bigg(H_{n}(\frac{\sqrt{2} x}{w_{0}}) \\
&+ \frac{\sqrt{2} x}{w_{0}} (\epsilon^2-\epsilon) 2 n H_{n-1}(\frac{\sqrt{2} x}{w_{0}}) + \frac{1}{2}\frac{2x^2}{w_{0}^2} (\epsilon^2-\epsilon)^2 4n(n-1)\\
&\times H_{n-2}(\frac{\sqrt{2} x}{w_{0}})\Bigg) e^{-\frac{x^2}{w_{0}^2}} \Big(1 + (2\epsilon - 3 \epsilon^2)\frac{x^2}{w_{0}^2} + \frac{2 x^4}{w_{0}^4} \epsilon^2 \Big) \,,
\end{aligned}
\end{equation} 
where we assume $\epsilon \ll 1$ and Taylor-expand to second order. Applying identities ~\ref{equ:hermite11} and~\ref{equ:hermite2} gives
\begin{equation}
\begin{aligned}
&\mathcal{U}_{\mathrm{n}}^{WS}(x, z) = \mathcal{U}_{\mathrm{n}} + \frac{\epsilon}{2}\Big(\sqrt{(n+1)(n+2)} \mathcal{U}_{\mathrm{n+2}} e^{-2\mathrm{i}\Psi} - \sqrt{n(n-1)} \\
&\times \mathcal{U}_{\mathrm{n-2}} e^{2\mathrm{i}\Psi}\Big) + \frac{\epsilon^2}{8} \bigg(
\sqrt{n(n-1)(n-2)(n-3)}\mathcal{U}_{\mathrm{n-4}}e^{4\mathrm{i}\Psi} \\
&+ 2\sqrt{n(n-1)}\mathcal{U}_{\mathrm{n-2}}e^{2\mathrm{i}\Psi} -2(n^2+n+1)\mathcal{U}_{\mathrm{n}} - 2\sqrt{(n+1)(n+2)}\\
&\times \mathcal{U}_{\mathrm{n+2}}e^{-2\mathrm{i}\Psi} + \sqrt{(n+1)(n+2)(n+3)(n+4)}\mathcal{U}_{\mathrm{n+4}}e^{-4\mathrm{i}\Psi}
\bigg)
\end{aligned}
\end{equation}

The mode coupling coefficient in the $x$-direction is then
\begin{equation}
\begin{aligned}
\rho_{x} &= \int_{-\infty}^{\infty} dx \cdot \mathcal{U}_{n}^{WS} \cdot \mathcal{U}_{\mathrm{n}}^{\ast}  \approx 1 - \frac{\epsilon^2}{4}\left(n^2+n+1\right)
\end{aligned}
\end{equation}
--note that unlike Eq.~\ref{equ:linear} there is no linear term in this case. We will have a similar result for $\rho_{y}$; the full coupling coefficient for $\mathrm{HG_{n,m}}$ due to waist size mismatch is therefore
\begin{equation}
\rho \approx 1 - \frac{\left(n^2+n+1\right)+\left(m^2+m+1\right)}{4} \cdot \epsilon^2
\end{equation}
The power coupling coefficient (Eq.~\ref{equ:genpowcoeff}) is thus
\begin{equation}
\eta^{WS} \approx 1 - \frac{\left(n^2+n+1\right)+\left(m^2+m+1\right)}{2} \cdot \epsilon^2\,,
\label{equ:powercoeff}
\end{equation}
and again we find a quadratic relationship for the relative power coupling loss (Eq.~\ref{equ:powerloss}) in the case of a waist size mismatch:
\begin{equation}
\Gamma_{n,m}^{WS} = \frac{n^2+n+m^2+m+2}{2}\,.
\label{equ:WSm}
\end{equation}

This matches the result for waist position mismatch, Eq.~\ref{equ:WPm}. Evaluation of Eqs.~\ref{equ:WPm} and~\ref{equ:WSm} exactly reproduces the coefficients found by A. Jones et. al.~\cite{Jones_2020}.

\section{Numerical comparison}
\label{sec:3}
\begin{figure}[htbp]
    \centering
    \includegraphics[width=\linewidth]{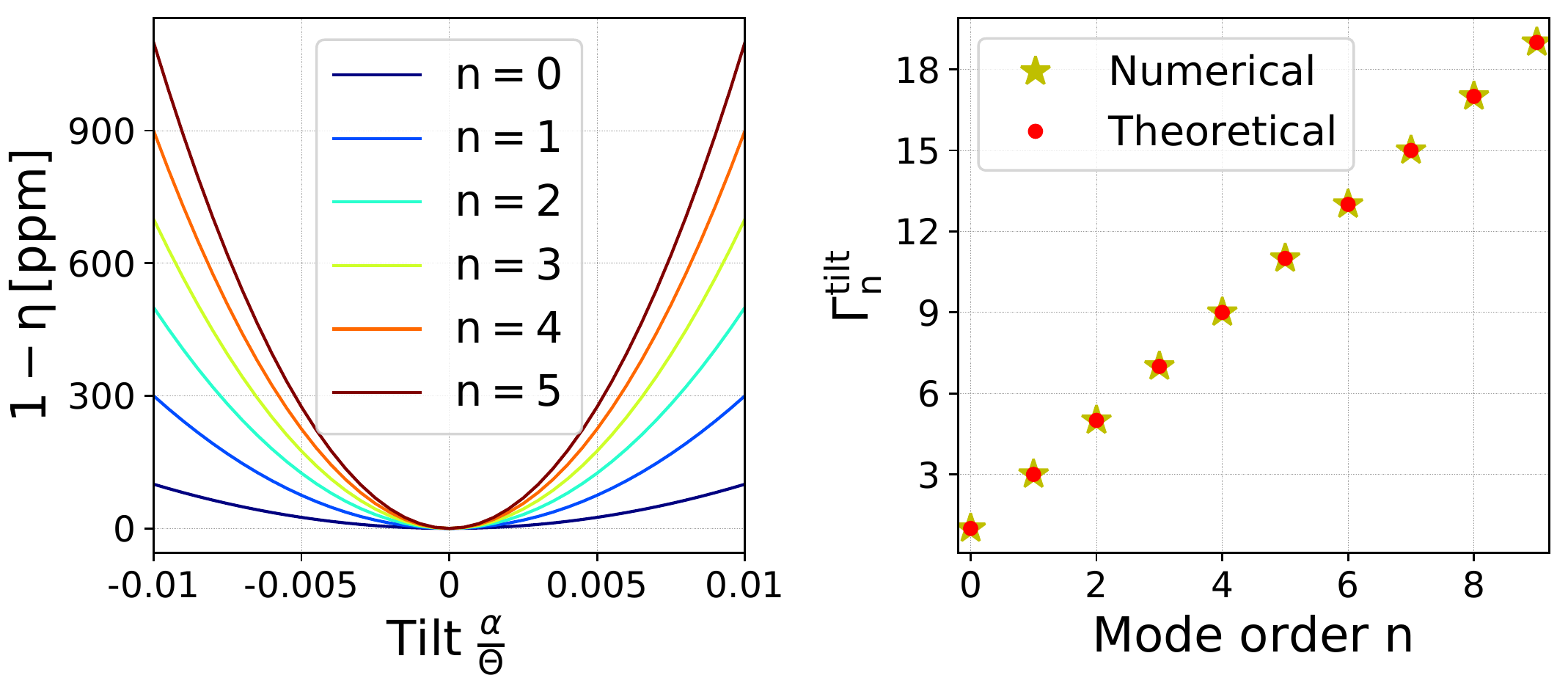}
    \caption{
    Left: Numerical power coupling loss in ppm as a function of tilt angle for $\mathrm{HG}_{\mathrm{n},0}$ modes; Right: Relative power coupling loss as a function of mode order.
    }
    \label{fig:powercouplings}
\end{figure}

The power coupling coefficients are calculated numerically by evaluating the overlap integrals (i.e. $\rho$) of discretized perturbed and original beams. Each beam is modeled as a 2-dimensional matrix of field amplitudes in the $x$-$y$ plane at the cavity waist; the integrals are evaluated using element-wise matrix multiplication. This is repeated for a range of perturbation amplitudes. The result for tilting $\mathrm{HG}_{\mathrm{n},0}$ modes is shown on the left of Fig.~\ref{fig:powercouplings}.

The relative power coupling loss (Eq.~\ref{equ:powerloss}) can be obtained numerically by taking the discretized second derivative of the overlap integral at zero perturbation. The normalised result for the case of tilted input is shown on the right of Fig.~\ref{fig:powercouplings}. The numerical result (yellow) agrees well with the analytical result (red) from Eq.~\ref{equ:tiltloss}. A similar result can be obtained for offsets.

\begin{figure}[htbp]
    \centering
    \includegraphics[width=\linewidth]{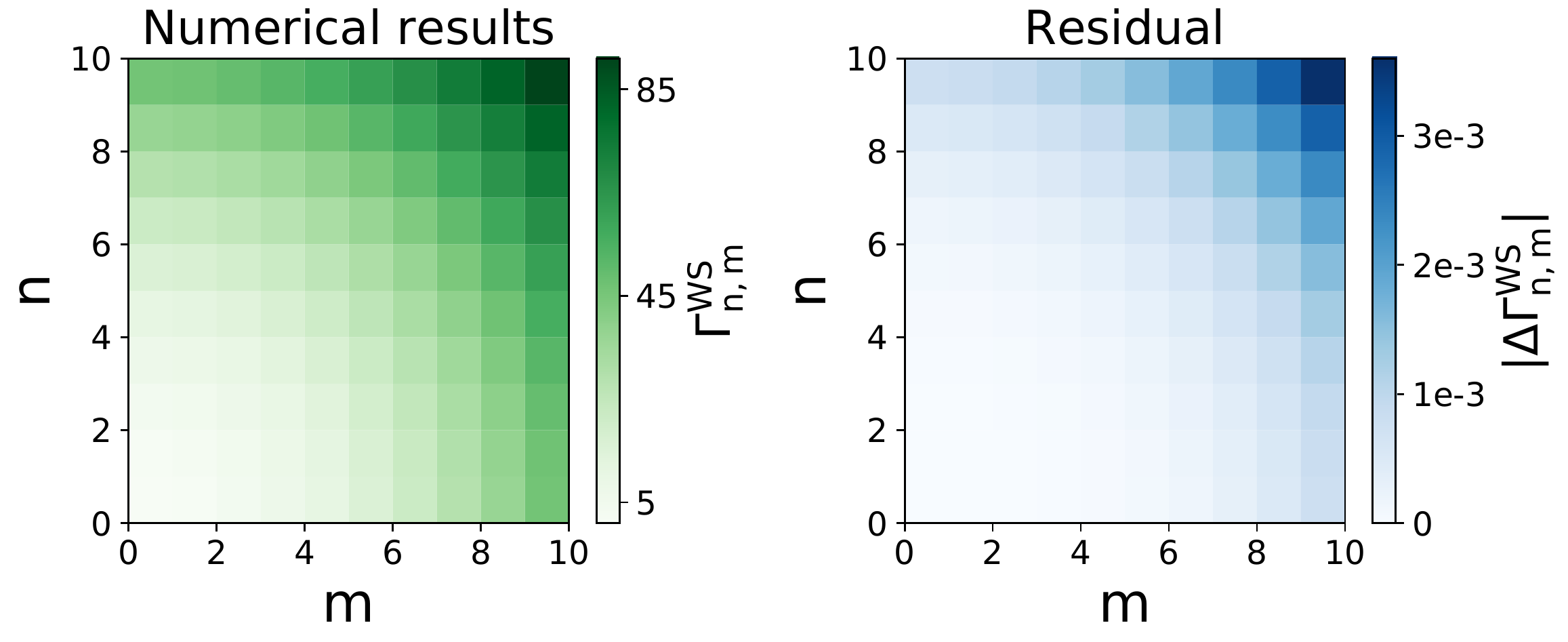}
    \caption{Left: numerical relative power coupling loss for $\mathrm{HG}_{\mathrm{n},\mathrm{m}}$ modes due to waist size mismatch; 
    Right: residuals when compared to Eq.~\ref{equ:WSm}.
    }
    \label{fig:powercouplinglossWS}
\end{figure}

The same method is used to calculate the power coupling coefficients and losses for mode mismatched $\mathrm{HG}_{\mathrm{n},\mathrm{m}}$ input beams. In this case the relative power coupling loss scales with both $n$ and $m$, as shown on the left of Fig.~\ref{fig:powercouplinglossWS} for waist size mismatches, becoming larger as we move up and right. The right panel of Fig.~\ref{fig:powercouplinglossWS} compares the numerical results to Eq.~\ref{equ:WSm}, in terms of the magnitude of the difference in the results $|\Delta\Gamma_\mathrm{n,m}^\mathrm{WS}|$. This residual is small, but increases with 
mode index
because higher-order modes with more high spatial frequency content need a finer grid resolution to match the accuracy of a lower-order model. A similar result can be obtained for waist position mismatches.

\section{Conclusion and Discussion}
\label{sec:4}
Through analytical and numerical methods we find that misalignment and mode-mismatch tolerances for higher-order HG mode beam are tighter than for $\mathrm{HG}_{0,0}$, as the induced relative power coupling losses scale linearly and quadratically with the mode order respectively. 

The maximum allowable mode mismatch for higher-order modes is smaller than for the fundamental mode, given the same mode-mismatch-induced power loss requirement. Specifically, since we are considering the second order expansion, using e.g. Eq.~\ref{equ:WSm} we see that in general the ratio of the maximum allowable mode mismatch for the $\mathrm{HG}_{\mathrm{n},\mathrm{m}}$ mode compared against the fundamental mode, given the same power loss requirement, is $\sqrt{\frac{2}{n^2+n+m^2+m+2}}$. For $\mathrm{HG}_{3,3}$ this is around 0.28.

Future work will investigate alignment and mode-matching sensing and control for arbitrary higher-order HG modes in various sensing schemes. This paper has shown that higher-order HG modes lead to tighter tolerances - if the same principals also lead to higher signal-to-noise in sensing schemes, as has been shown for Laguerre-Gauss modes~\cite{PhysRevD.79.122002}, this will help mitigate the challenges discussed here.

\section{Funding}
This work was supported by National Science Foundation grants PHY-1806461 and PHY-2012021.

\bibliographystyle{unsrt}
\bibliography{template}

\begin{thebibliography}{10}

\bibitem{aLIGO}
J.~Aasi et. al.
\newblock Advanced {LIGO}.
\newblock {\em Classical and Quantum Gravity}, 32(7):074001, mar 2015.

\bibitem{AdVirgo}
F.~Acernese et. al.
\newblock Advanced virgo: a second-generation interferometric gravitational
  wave detector.
\newblock {\em Classical and Quantum Gravity}, 32(2):024001, dec 2014.

\bibitem{Mours_2006}
Beno{\^{\i}}t Mours, Edwige Tournefier, and Jean-Yves Vinet.
\newblock Thermal noise reduction in interferometric gravitational wave
  antennas: using high order {TEM} modes.
\newblock {\em Classical and Quantum Gravity}, 23(20):5777--5784, sep 2006.

\bibitem{Vinet_2007}
Jean-Yves Vinet.
\newblock Reducing thermal effects in mirrors of advanced gravitational wave
  interferometric detectors.
\newblock {\em Classical and Quantum Gravity}, 24(15):3897--3910, jul 2007.

\bibitem{PhysRevD.102.122002}
Liu Tao, Anna Green, and Paul Fulda.
\newblock Higher-order hermite-gauss modes as a robust flat beam in
  interferometric gravitational wave detectors.
\newblock {\em Phys. Rev. D}, 102:122002, Dec 2020.

\bibitem{PhysRevD.103.042008}
Stefan Ast, Sibilla Di~Pace, Jacques Millo, Mikha\"el Pichot, Margherita
  Turconi, Nelson Christensen, and Walid Chaibi.
\newblock Higher-order hermite-gauss modes for gravitational waves detection.
\newblock {\em Phys. Rev. D}, 103:042008, Feb 2021.

\bibitem{Jones_2020}
A.~W. Jones and A.~Freise.
\newblock Increased sensitivity of higher-order laser beams to mode mismatches.
\newblock {\em Optics Letters}, 45(20):5876, Oct 2020.

\bibitem{sympy}
Meurer A, Paprocki~M Smith~CP, Kirpichev~SB Čertík O, Kumar~A Rocklin~M,
  Moore~JK Ivanov~S, Rathnayake~T Singh~S, Granger~BE Vig~S, Bonazzi~F
  Muller~RP, Vats~S Gupta~H, Pedregosa~F Johansson~F, Terrel~AR Curry~MJ,
  Saboo~A Roučka~Š, Kulal~S Fernando~I, Cimrman R, and Scopatz A.
\newblock Sympy: symbolic computing in python, jan 2017.
\newblock PeerJ Computer Science 3:e103
  \url{https://doi.org/10.7717/peerj-cs.103}.

\bibitem{Bayer-Helms:84}
F.~Bayer-Helms.
\newblock Coupling coefficients of an incident wave and the modes of a
  spherical optical resonator in the case of mismatching and misalignment.
\newblock {\em Appl. Opt.}, 23(9):1369--1380, May 1984.

\bibitem{Bond2017}
Charlotte Bond, Daniel Brown, Andreas Freise, and Kenneth~A Strain.
\newblock Interferometer techniques for gravitational-wave detection.
\newblock {\em Living Reviews in Relativity}, 19, Feb 2017.

\bibitem{Morrison1:94}
Euan Morrison, Brian~J. Meers, David~I. Robertson, and Henry Ward.
\newblock Automatic alignment of optical interferometers.
\newblock {\em Appl. Opt.}, 33(22):5041--5049, Aug 1994.

\bibitem{PhysRevD.79.122002}
Simon Chelkowski, Stefan Hild, and Andreas Freise.
\newblock Prospects of higher-order laguerre-gauss modes in future
  gravitational wave detectors.
\newblock {\em Phys. Rev. D}, 79:122002, Jun 2009.

\end{thebibliography}

\end{document}